\documentclass[12pt]{article}

\usepackage{graphicx}
\usepackage{a4}

\usepackage{fancybox}
\usepackage{epsfig}
\usepackage{color}

\usepackage{amsfonts}
\usepackage{mathrsfs}
\usepackage{amssymb}
\usepackage{amsmath}

\usepackage{dcolumn}
\usepackage{bm}

\def\pa{\partial}

\def\al{\alpha}
\def\be{\beta}
\def\ga{\gamma}
\def\de{\delta}
\def\ep{\epsilon}

\def\th{\theta}

\def\la{\lambda}

\def\si{\sigma}

\def\om{\omega}

\def\Ga{\Gamma}

\def\La{\Lambda}

\def\Om{\Omega}

\newcommand{\ben}{\begin{equation}}
\newcommand{\een}{\end{equation}}
\newcommand{\bea}{\begin{eqnarray}}
\newcommand{\eea}{\end{eqnarray}}
\newcommand{\ba}{\begin{array}}
\newcommand{\ea}{\end{array}}
\newcommand{\bit}{\begin{itemize}}
\newcommand{\eit}{\end{itemize}}

\newcommand{\vs}[1]{\vspace{#1 mm}}

\newcommand{\dsl}{\pa \kern-0.5em /}

\begin{document}

\topmargin 0pt \oddsidemargin 0mm

\begin{flushright}

USTC-ICTS-11-10 \\

\end{flushright}

\vspace{2mm}

\begin{center}

{\Large \bf AdS segment and hidden conformal symmetry in general
non-extremal black holes}

\vs{10}

 {\large Huiquan Li \footnote{E-mail: hqli@ustc.edu.cn}}

\vspace{6mm}

{\em

Interdisciplinary Center for Theoretical Study\\

University of Science and Technology of China, Hefei, Anhui 230026, China\\

}

\end{center}

\vs{9}

\begin{abstract}
It is demonstrated that the near-horizon geometry of general
non-extremal black holes can be described by a portion of AdS space.
We show that the reason why hidden conformal symmetries near
horizons of general non-extremal black holes are achieved in
previous works is that the near-horizon geometries have been
equivalently taken as these AdS segments rather than simply the
Rindler space.
\end{abstract}



\section{Introduction}
\label{sec:introduction}
%%%%%%%%%%%%%%%%%%%%%%%%%%%%%%%%%%%%%%%%%%%%%%%%%%%%%%%%%%%%%%%%%%%%%%%%%%%%

According to the AdS/CFT correspondence, the AdS structure
\cite{Bardeen:1999px} near horizons of extreme Kerr black holes
implies that Kerr black holes at extremality could be described by a
two-dimensional conformal field theory (CFT) \cite{Guica:2008mu}
(for a review see \cite{Bredberg:2011hp}). The near-horizon geometry
of extreme Kerr is a warped $AdS_3$ space with the $SL(2,R)\times
U(1)$ isometry group. Through the evaluation of the asymptotic
behavior by choosing appropriate boundary conditions, it is revealed
that the asymptotic symmetry generators form one copy of Virasoro
algebra with central charge $c_L=12J$, enhanced from the $U(1)$
isometry group \cite{Guica:2008mu}. It is suggested that extreme
Kerr black holes should be holographically dual to a chiral CFT with
the dimensionless temperature $T_L=1/2\pi$. This is supported by the
success in recovering the Bekenstein-Hawking (BH) entropy of extreme
Kerr black holes using the dual CFT. In the near-extreme case, the
near-horizon geometry is asymptotically the $AdS_3$ space of extreme
Kerr and a second copy of Virasoro algebra with the same central
charge $c_R=12J$ emerges, being an enhancement from the $SL(2,R)$
isometry \cite{Matsuo:2009sj,Rasmussen:2009ix,Castro:2009jf}. This
corresponds to a two-dimensional CFT with opposite chirality.

It is hard to go beyond linear deviations from extremality. In this
case, we need to consider effects of backreaction
\cite{Maldacena:1998uz,Castro:2010fd}. The near-horizon geometry of
off-extreme kerr is Rindler space in leading order and is no longer
asymptotically a warped $AdS_3$ space. Hence, no conformal symmetry
is expected in the Kerr black holes far away from extremality.
However, an $SL(2,R)_L\times SL(2,R)_R$ symmetry is still revealed
in the Klein-Gordon equation of a minimally coupled massless scalar
propagating in ``near regions" of generic non-extreme Kerr with low
frequencies \cite{Castro:2010fd}. The $SL(2,R)_L\times SL(2,R)_R$
symmetry in the wave equation spontaneously breaks down to
$U(1)\times U(1)$ by the temperatures $T_L=M^2/(2\pi J)$ and
$T_R=\sqrt{M^4-J^2}/(2\pi J)$.  This conformal symmetry is ``hidden"
because there is no geometry accounting for it in off-extreme case.
It implies that the general Kerr should be dual to a two-dimensioanl
CFT with central charges $c_L=c_R=12J$ at temperatures $(T_L,T_R)$.
The entropy of general Kerr black holes computed by using Cardy's
formula agrees well with the BH area law for black hole entropy and
the result appears to be valid for any values of $M$ and $J$.

More hidden conformal symmetries are found in various black hole
spacetimes
\cite{Chen:2010as,Chen:2010yu,Wang:2010qv,Chen:2010xu,Chen:2010zwa,
Wang:2010ic,Shao:2010cf,Setare:2010sy,Ghezelbash:2010vt,Huang:2010yg,
Peng:2011zza,Chen:2011kt}. Recently, it is claimed in
\cite{Bertini:2011ga} that there is also an $SL(2,R)$ symmetry in
the wave equation in near regions of Schwarzschild black holes.
Actually, the conformal symmetry near horizons of general black
holes, including Schwarzschild black holes, has long been suggested
from the analysis on diffeomorphisms
\cite{Carlip:1998wz,Solodukhin:1998tc} and via the relation between
optical metric and AdS space \cite{Sachs:2001qb}. Further evidence
about this is given in
\cite{Govindarajan:2000ag,Birmingham:2001qa,Gupta:2001bg}, which
shows that the near-horizon dynamics of the Klein-Gordon equation
for a scalar is a De Alfaro-Fubini-Furlan (DDF) model of conformal
quantum mechanics \cite{deAlfaro:1976je}.

Various attempts to understand the hidden conformal symmetries have
been made in \cite{Matsuo:2010in,Cvetic:2011hp,Franzin:2011wi}. In
the present work, we argue that, in derivation of hidden conformal
symmetries near horizons in previous works, the near-horizon
geometries of non-extreme black holes are actually taken as some
particular AdS segments instead of simply Rindler space. The AdS
segment shrinks to Rindler space in off-extreme case and grows to
the AdS space that is usually derived in near-extreme case.

The paper is organised as follows. In Section
\ref{sec:adsdescription}, we show that the near-horizon geometries
of general non-extreme black holes can be described by AdS segments.
In sections \ref{sec:ads2} and \ref{sec:ads3}, we show how these AdS
segments rather than Rindler space are responsible for the hidden
conformal symmetries near horizons in Scharzschild/RN and Kerr black
holes, respectively. We give conclusions in the final section. The
Appendix presents the AdS segment near horizons of 5D rotating black
holes.

%%%%%%%%%%%%%%%%%%%%%%%%%%%%%%%%%%%%%%%%%%%%%%%%%%%%%%%%%%%%%%%%%%%%%%%%%%%%
\section{AdS segments in non-extremal black holes}
\label{sec:adsdescription}
%%%%%%%%%%%%%%%%%%%%%%%%%%%%%%%%%%%%%%%%%%%%%%%%%%%%%%%%%%%%%%%%%%%%%%%%%%%%

In this section, we show that the near-horizon geometry of general
non-extremal black holes can be described by a portion of AdS space.
For black holes away from extremality, these AdS segments are
undistinguishable from Rindler space in leading order. Towards the
extreme limit, they can continuously grow to the AdS spaces that are
previously known in near-extreme case. In later sections, we shall
show that it is these AdS spaces that are responsible for the
``hidden" conformal symmetries.

\subsection{Schwarzschild black holes}

Let us start with the Schwarzschild metric
\begin{equation}
 ds^2=-\left(1-\frac{r_0}{r}\right)dt^2+\left(1-\frac{r_0}{r}
\right)^{-1}dr^2+r^2(d\th^2+\sin^2\th d\phi^2),
\end{equation}
where $r_0=2M$ and $M$ is the mass of the black hole. It is known
that the near horizon geometry of the Schwarzschild black hole is a
product of the Rindler space and a two sphere $S^2$, which is, with
the coordinate redefinitions $r=r_0(1+\rho^2/4)$ and
$\hat{t}=t/(2r_0)$
\begin{equation}\label{e:rindler}
 ds^2=r_0^2(-\rho^2d\hat{t}^2+d\rho^2+d\Om_2^2).
\textrm{ }\textrm{ }\textrm{ } (0\leq\rho\ll1)
\end{equation}
The Rindler space describes a uniformly accelerating frame.

In this paper, we show that the reason why hidden conformal symmetry
is achieved in Schwarzschild spacetime in \cite{Bertini:2011ga} is
that the authors have essentially adopted the following $AdS_2$
segment instead of the Rindler space (\ref{e:rindler}) as the
near-horizon geometry
\begin{equation}\label{e:rindlerads}
 ds^2=r_0^2(-\sinh^2\rho d\hat{t}^2+d\rho^2+d\Om_2^2).
\textrm{ }\textrm{ }\textrm{ } (0\leq\rho\ll1)
\end{equation}
This AdS space is undistinguishable from Rindler space in leading
order (actually in up to quadratic order) as the coordinate $\rho$
is small. But the two geometries accommodate different symmetries:
the $AdS_2$ space has an $SL(2,R)$ isometry group, while the Rindler
space does not. In Section \ref{sec:ads2}, we shall argue that the
hidden conformal symmetry observed in Schwarzschild spacetime
actually arises from this $AdS_2$ segment. If the near-horizon
geometry is exactly the Rindler space, there is no such conformal
symmetry. In the next subsection, we shall show why the near-horizon
geometry of Schwarzschild black hole can be described by the AdS
segment (\ref{e:rindlerads}), by analysing the Reissner-Nordstr{\o}m
(RN) metric.

\subsection{The AdS$_2$ segment from RN metric}

It is clearer to see the origin of the AdS segment
(\ref{e:rindlerads}) in the RN metric for a charged black hole:
\begin{equation}\label{e:RN}
 ds^2=-\frac{(r-r_+)(r-r_-)}{r^2}dt^2+\frac{r^2}{(r-r_+)(r-r_-)}
dr^2+r^2d\Om_2^2.
\end{equation}
where $r_\pm$ denotes the radii of the inner and outer horizons,
respectively, which depend on the mass and charge of the black hole.
For the $r_-=0$ case, we go back to the Schwarzschild metric. For
the $r_+=r_-$ case, we get the extreme limit.

It is convenient to redefine the coordinates in the following way
\begin{equation}\label{e:RNred1}
 r=r_+(1+\de), \textrm{ }\textrm{ }\textrm{ }
\de=\ep\sinh^2\frac{\rho}{2}, \textrm{ }\textrm{ }\textrm{ }
\ep=\frac{r_+-r_-}{r_+}, \textrm{ }\textrm{ }\textrm{ }
t=\frac{2r_+}{\ep}\hat{t}.
\end{equation}
Then, the metric in Eq.\ (\ref{e:RN}) is exactly rewritten as
\begin{equation}\label{e:rnrewrite}
 ds^2=r_+^2\left[-\frac{\sinh^2\rho}{(1+\de)^2}d\hat{t}^2+
(1+\de)^2d\rho^2+(1+\de)^2d\Om_2^2\right].
\end{equation}
In regions closed to the outer horizon at $r=r_+$, i.e.,
\begin{equation}\label{e:nearhorcond}
 \de=\ep\sinh^2\frac{\rho}{2}\ll1,
\end{equation}
we approximately get a portion of $AdS_2\times S^2$:
\begin{equation}\label{e:rnrewrite1}
 ds^2=r_+^2(-\sinh^2\rho d\hat{t}^2+
d\rho^2+d\Om_2^2), \textrm{ }\textrm{ }\textrm{ }
0\leq\rho\ll2\sinh^{-1}\frac{1}{\sqrt{\ep}}.
\end{equation}
The cutoff of the coordinate $\rho$ is constrained by a function of
the parameter $\ep$, which measures how extremal the black hole is.
It is seen that, far away from extremality: $\ep\sim1$ ($\ep=1$ is
the Schwarzschild case), the valid value of $\rho$ must be very
small $0\leq\rho\ll1$. For the $\ep=1$ case, we get the AdS$_2$
segment (\ref{e:rindlerads}) for Schwarzschild black hole, which is
undistinguishable from the Rindler space in leading order since
$\sinh\rho\simeq\rho$. In the extreme limit $\ep\rightarrow0$, the
maximum value of $\rho$ now can be very large: $\rho\in[0,\infty)$,
as the condition (\ref{e:nearhorcond}) is still saturated. It is
easy to prove that the part insider the horizon is also a portion of
AdS space, as in \cite{Carroll:2009maa}.

There is a subtlety for the extreme limit of the RN black hole. The
non-extreme RN metric splits into two spacetimes at extremality: the
disconnected AdS$_2$ space and the extremal black hole
\cite{Carroll:2009maa,Garousi:2009zx}. The former outside the outer
horizon takes the form (\ref{e:rnrewrite1}) with
$\rho\in[0,\infty)$. The extremal RN metric takes the form:
\begin{equation}\label{e:extads}
 ds_2^2=R^2\left(-u^2{d\tau}^2+\frac{du^2}{u^2}
\right),
\end{equation}
with $R=r_+$, which is the near-horizon geometry for points with a
finite distance away from the horizon in the extremal limit. Both
AdS$_2$ metric are patches of the fully extended AdS$_2$ space. They
can be related via the following relations
\begin{equation}\label{e:adsmapsol}
 u=\frac{1}{R}\sinh\rho e^{-\hat{t}},
\textrm{ }\textrm{ }\textrm{ } \tau=R\coth\rho e^{\hat{t}}.
\end{equation}
Thus, whatever the value of $\rho$ is, the AdS$_2$ space
(\ref{e:rnrewrite1}) will always evolve into the horizon (where
$u=0$) of the extremal black hole solution (\ref{e:extads}) as
$\hat{t}\rightarrow\infty$. This implies that there should be the
extremal Kerr geometry (\ref{e:extads}) left eventually.

\subsection{Kerr black holes}

It is straightforward to make the similar analysis in the Kerr
metric
\begin{equation}\label{e:Kerr}
 ds^2=-\frac{\triangle}{\chi^2}(dt-a\sin^2\th d\phi)^2+\frac{\sin^2\th}
{\chi^2}((r^2+a^2)d\phi-adt)^2+\frac{\chi^2}{\triangle}dr^2+\chi^2d\th^2,
\end{equation}
where
\begin{equation}
 \triangle=(r-r_+)(r-r_-), \textrm{ }\textrm{ }\textrm{ }
r_\pm=M\pm\sqrt{M^2-a^2} \nonumber
\end{equation}
\begin{equation}
 \chi^2=r^2+a^2\cos^2\th, \textrm{ }\textrm{ }\textrm{ }
a=\frac{J}{M}. \nonumber
\end{equation}
$r_\pm$ are the radii of the inner and outer horizons, respectively,
and $J$ is the angular momentum of the black hole.

We adopt the following notations and coordinate redefinitions:
\begin{equation}\label{e:Kerrred1}
 r=r_+(1+\de), \textrm{ }\textrm{ }\textrm{ }
\de=\la U, \textrm{ }\textrm{ }\textrm{ } t=\frac{r_+\bar{t}}{\la},
\end{equation}
\begin{equation}\label{e:Kerrred2}
 \phi=\bar{\phi}+\frac{\al t}{r_+(1+\al^2)},  \textrm{ }\textrm{ }\textrm{ }
\al=\frac{a}{r_+}, \textrm{ }\textrm{ }\textrm{ }
\la\ep=\frac{r_+-r_-}{r_+}=1-\al^2.
\end{equation}
The last equation comes from the relation: $r_+r_-=a^2$. Then the
Kerr metric (\ref{e:Kerr}) can be exactly rewritten as
\begin{eqnarray}\label{e:genKerr}
 ds^2=r_+^2\left\{-\frac{U(\ep+U)}{x^2}\left(\frac{1+\al^2
\cos^2\th}{1+\al^2}d\bar{t}-\la\al\sin^2\th
d\bar{\phi}\right)^2+\frac{x^2}{U(\ep+U)}dU^2 \right. \nonumber \\
\left.+\frac{\sin^2\th}{x^2}\left[((1+\de)^2+\al^2)d\bar{\phi}
+\frac{U(2+\de)}{1+\al^2}\al d\bar{t}\right]^2+x^2d\th^2\right\},
\end{eqnarray}
where
\begin{equation}
 x^2=\frac{\chi^2}{r_+^2}=(1+\de)^2+\al^2\cos^2\th.
\end{equation}

Taking the limits
\begin{equation}\label{e:KerrNHcon}
 \la\ll1 \textrm{ }\textrm{ }\textrm{ and }\textrm{ }\textrm{ }
\de=\la U\ll1,
\end{equation}
and redefining
\begin{equation}\label{e:Kerrred3}
 \bar{t}=(1+\al^2)\widetilde{t}, \textrm{ }\textrm{ }\textrm{ }
\bar{\phi}=\frac{2\al\widetilde{\phi}}{1+\al^2},
\end{equation}
we can get the near-horizon geometry:
\begin{eqnarray}\label{e:genKerrAdS1}
 ds^2=r_+^2\Ga\left\{-U(\ep+U)d\widetilde{t}^2+\frac{dU^2}{U(\ep+U)}
+d\th^2+\frac{4\al^2\sin^2\th}{\Ga^2}(d\widetilde{\phi}+
Ud\widetilde{t})^2\right\},
\end{eqnarray}
where
\begin{equation}\label{e:Ga}
  \Ga(\th)=1+\al^2\cos^2\th.
\end{equation}
Note that $\ep$ here satisfies the following condition, from the
constraint (\ref{e:KerrNHcon})
\begin{equation}\label{e:Kerrepcon}
 \ep=\frac{1-\al^2}{\la}\gg1-\al^2.
\end{equation}
Far away from extremality $\al\rightarrow0$, $\ep$ can only be very
large values $\ep\gg1$. Towards the extreme limit $\al\rightarrow1$,
$\ep$ is possible to be small values. The conditions in Eq.\
(\ref{e:KerrNHcon}) also require that $U$ should not be too large.
Without loss of generality, we may set $U$ to be order of unity:
$U\sim\mathcal{O}(1)$ in the following discussion.

Further doing the coordinate redefinitions
\begin{equation}\label{e:Kerrred4}
  \widetilde{t}=\frac{2\hat{t}}{\ep},
\textrm{ }\textrm{ }\textrm{ } \widetilde{\phi}=\hat{\phi}, \textrm{
}\textrm{ }\textrm{ }\textrm{ }\textrm{ }\textrm{ }
U=\ep\sinh^2\frac{\rho}{2}.
\end{equation}
the near-horizon geometry (\ref{e:genKerrAdS1}) for a general Kerr
black hole is reexpressed as
\begin{equation}\label{e:genKerrAdS2}
 ds^2=r_+^2\Ga\left\{-\sinh^2\rho d\hat{t}^2+d\rho^2+
d\th^2+\frac{4\al^2\sin^2\th}{\Ga^2}[d\hat{\phi}+(\cosh\rho
-1)d\hat{t}]^2\right\}.
\end{equation}
where
\begin{equation}\label{e:KerrAdSrhorang}
 \rho\sim2\sinh^{-1}\left(\frac{1}{\sqrt{\ep}}\right)
\ll2\sinh^{-1}\left(\frac{1}{\sqrt{1-\al^2}}\right).
\end{equation}
This geometry looks like a portion of warped AdS$_3$ space, with the
coordinate $\rho$ varying within a range constrained by Eq.\
(\ref{e:KerrAdSrhorang}). Only in the extreme limit
$\ep\rightarrow0$ or $\al\rightarrow1$, $\rho$ can reach infinity.
Far away from extremality $\ep\rightarrow1$, this geometry is
undistinguishable from the Rindler space in leading order since
$\rho$ can only be small values due to the constraint
(\ref{e:KerrAdSrhorang}): $\sinh\rho\sim\rho$ and $\cosh\rho\sim1$.

An alternative form of the near-horizon geometry can be obtained via
the following redefinitions of the $(t,\phi)$ coordinates (keeping
other notations and redefinitions the same):
\begin{equation}\label{e:Kerrredal}
 t=\frac{4r_+}{1-\al^2}\hat{t}, \textrm{ }\textrm{ }\textrm{ }
\phi=\frac{2\hat{t}}{\al(1-\al^2)}+
\frac{2\hat{\phi}}{\al(1+\al^2)}.
\end{equation}
The metric becomes by taking the near-horizon limits
(\ref{e:KerrNHcon})
\begin{equation}\label{e:genKerrAdS3}
  ds^2=r_+^2\Ga\left[-\sinh^2\rho\frac{(1+\cos^2\th)^2}{\Ga^2}
d\hat{t}^2+d\rho^2+d\th^2+\frac{4\sin^2\th}
{\al^2\Ga^2}(d\hat{\phi}+\cosh\rho d\hat{t})^2\right],
\end{equation}
where $\Ga(\th)$ is the same as defined in Eq.\ (\ref{e:Ga}) and
$\rho$ satisfies the condition (\ref{e:KerrAdSrhorang}) as well.
This is more like a portion of warped AdS$_3$ space. Towards the
extreme limit, $\rho$ can reach infinity. In this limit, we have
$\Ga\rightarrow1+\cos^2\th$ and so recover the warped AdS$_3$ space
obtained in \cite{Castro:2009jf} in the near-extreme case.

In Section \ref{sec:ads3}, we shall show that a universal $AdS_3$
segment like (\ref{e:genKerrAdS2}) or (\ref{e:genKerrAdS3}) in
general non-extremal case should be responsible for the near-horizon
hidden conformal symmetry \cite{Castro:2010fd} observed in wave
equation in non-extremal Kerr spacetime.

%%%%%%%%%%%%%%%%%%%%%%%%%%%%%%%%%%%%%%%%%%%%%%%%%%%%%%%%%%%%%%%%%%%%%%%%%%%%
\section{Conformal symmetry in general RN/Schwarzschild}
\label{sec:ads2}
%%%%%%%%%%%%%%%%%%%%%%%%%%%%%%%%%%%%%%%%%%%%%%%%%%%%%%%%%%%%%%%%%%%%%%%%%%%%

In this section, we show that it is the AdS$_2$ segment
(\ref{e:rindlerads}) or (\ref{e:rnrewrite1}) that accounts for the
near-horizon ``hidden" conformal symmetries in Schwarzschild and
non-extremal RN black holes. The hidden conformal symmetry is
observed in the wave equation of a massless scalar propagating in
the corresponding black hole backgrounds. Specifically, we can
consider the following Klein-Gordon equation
\begin{equation}\label{e:KGeq}
 \frac{1}{\sqrt{-g}}\pa_\mu(\sqrt{-g}g^{\mu\nu}\pa_\nu\Phi)=0.
\end{equation}

For the RN metric (including the Schwarzschild case), the equation
is
\begin{equation}\label{e:RNKGeq}
 \left[-\frac{r^4}{\triangle}\pa_t^2+\pa_r\triangle\pa_r+
\nabla_{S^2}^2\right]\Phi=0,
\end{equation}
where $\triangle=(r-r_-)(r-r_+)$ and
\begin{equation}\label{e:S2}
 \nabla_{S^2}^2=\frac{1}{\sin\th}\pa_\th\sin\th
\pa_\th+\frac{1}{\sin^2\th}\pa_\phi^2.
\end{equation}
We can do the separation:
\begin{equation}\label{e:RNscalar}
 \Phi=R(r,t)Y_m^l(\th,\phi),
\end{equation}
where $Y_m^l(\th)$ ($|m|\leq l$) are the spherical harmonic
solutions. Eq.\ (\ref{e:RNKGeq}) can be decomposed into the two
equations
\begin{equation}\label{e:}
 \nabla_{S^2}^2Y_m^l(\th,\phi)=-l(l+1)
Y_m^l(\th,\phi),
\end{equation}
\begin{equation}\label{e:RNradialeq}
 \left[\pa_r\triangle\pa_r-\frac{r^4}{\triangle}\pa_t^2-l(l+1)
\right]R(r,t)=0.
\end{equation}

As done in \cite{Bertini:2011ga}, under the ``near-region" condition
$\om r\le1$ and the low-frequency condition $\om r_+\le1$, the
radial equation becomes
\begin{equation}\label{e:origconfeq}
 \left[\pa_r\triangle\pa_r-\frac{r_+^4}{\triangle}\pa_t^2-l(l+1)
\right]R(r,t)=0.
\end{equation}
This equation contains solutions that are $SL(2,R)$ invariant.
Because the geometrical origin of this symmetry is unknown, it is
called ``hidden".

Now we show that the hidden conformal symmetry is derived because
the near-horizon geometry of general non-extreme RN has been taken
as the AdS$_2$ segment (\ref{e:rnrewrite1}). To do this, let us
adopt the coordinate redefinitions given in Eq.\ (\ref{e:RNred1})
that lead to the AdS$_2$ geometry (\ref{e:rnrewrite1}). In this
coordinate system, the scalar field (\ref{e:RNscalar}) is expressed
as: $\Phi=R(\rho,\hat{t})Y_m^l(\th,\phi)$, and the radial part of
the KG equation (\ref{e:RNKGeq}) is exactly rewritten as
\begin{equation}\label{e:RNredradeq}
 \left[\frac{1}{\sinh\rho}\pa_\rho\sinh\rho\pa_\rho+\frac{
(1+\de)^4}{\sinh^2\rho}\pa_{\hat{t}}^2-l(l+1)\right]
R(\rho,\hat{t})=0.
\end{equation}
Taking the near-horizon limit (\ref{e:nearhorcond}): $\de\ll1$, we
get the equation comparable to Eq.\ (\ref{e:origconfeq}), leading to
the $SL(2,R)$ symmetry. Note that, to get this equation, here we
adopt the near-horizon limit rather than the near-region and
low-frequency conditions. It can be proved that this is the KG
equation in the AdS space (\ref{e:rnrewrite1}). We can write down
the vectors
\begin{eqnarray}\label{e:chih}
 H_1(\rho,\hat{t})=ie^{\hat{t}}(\pa_\rho
-\coth\rho\pa_{\hat{t}}),
\nonumber \\
H_{-1}(\rho,\hat{t})=-ie^{-\hat{t}}
(\pa_\rho+\coth\rho\pa_{\hat{t}}),
\\
H_0(\rho,\hat{t})=-i\pa_{\hat{t}}, \nonumber
\end{eqnarray}
where $\rho$ satisfies the constraint (\ref{e:nearhorcond}). If we
use the coordinate redefinition relations (\ref{e:RNred1}) for the
Schwarzschild metric case, these vectors are exactly those obtained
in \cite{Bertini:2011ga}.

The vectors obey the $SO(2,1)$ or $SL(2,R)$ algebra:
\begin{eqnarray}
\label{e:hcom}
 [H_0,H_{\pm 1}]=\mp iH_{\pm 1}, & [H_1,H_{-1}]=2iH_0.
\end{eqnarray}
Eq.\ (\ref{e:RNredradeq}) in the near-horizon limit can be expressed
as the Casimir of the vectors:
$H^2R(\rho,\hat{t})=h(h-1)R(\rho,\hat{t})$, where the conformal
weight $h=l+1$ following the notation
\cite{Lowe:2011wu,Bertini:2011ga}. %$L_n=-iH_n$ ($n=-1,0,1$).
Since the generators and solution $\Phi$ are the same as derived in
\cite{Bertini:2011ga} (expressed in different coordinates), we can
also get the same relation $M\om_n=M\om_0-in/4$ for the descendents.

If the near-horizon geometry is Rindler space, the corresponding
wave equation to Eq.\ (\ref{e:RNradialeq}) becomes:
$[(1/\rho)\pa_\rho\rho\pa_\rho+\om^2/\rho^2]R(\rho)=l(l+1)R(\rho)$
by using the field redefinition relation $r=r_+(1+\rho^2/4)$. The
resulting vectors are those (\ref{e:chih}) with the replacement of
$\coth\rho\rightarrow1/\rho$. In this case, the first equation in
Eq.\ (\ref{e:hcom}) is still correct but the second one is not.
Thus, the $SL(2,R)$ algebra does not close. Hence, we can say that,
in derivation of the hidden conformal symmetry in
\cite{Bertini:2011ga}, the relevant near-horizon geometry of the
Schwarzschild or generic non-extreme RN black holes is not simply
the Rindler space any more, but the $AdS_2$ space
(\ref{e:rnrewrite1}).

Finally, we note that this near-horizon geometry can not provide
apparent explanation to the DDF model of the near-horizon conformal
dynamics \cite{Govindarajan:2000ag,Birmingham:2001qa,Gupta:2001bg}.
Using the AdS$_2$ segment (\ref{e:rnrewrite1}) and redefining
$R=\psi/\sqrt{\sinh\rho}$, we can approximately have the radial
equation
\begin{equation}\label{e:DDFeq}
 \frac{d^2\psi}{d\rho^2}+\frac{g\psi}{\rho^2}=0,
\end{equation}
where $g=1+(4M\om)^2$. Based on this Hamiltonian, generators obeying
the $SL(2,R)$ algebra can be constructed. However, using the Rindler
space instead, we can get the exact form of the equation
(\ref{e:DDFeq}) by redefining $R=\psi/\sqrt{\rho}$.

%%%%%%%%%%%%%%%%%%%%%%%%%%%%%%%%%%%%%%%%%%%%%%%%%%%%%%%%%%%%%%%%%%%%%%%%%%%%
\section{Conformal symmetry in general Kerr}
\label{sec:ads3}
%%%%%%%%%%%%%%%%%%%%%%%%%%%%%%%%%%%%%%%%%%%%%%%%%%%%%%%%%%%%%%%%%%%%%%%%%%%%

We now turn to the Kerr black holes. Following \cite{Castro:2010fd},
we separate the massless scalar as $\Phi=e^{-i\om
t}R(r)S(\th,\phi)$. In the Kerr spacetime, the Klein-Gordon equation
(\ref{e:KGeq}) decomposes into:
\begin{equation}\label{e:KerrKGeq1}
 (\nabla^2_{S^2}+\om^2a^2\cos^2\th)S(\th,\phi)=-K_lS(\th,\phi),
\end{equation}
\begin{equation}\label{e:KerrKGeq2}
 \left[\pa_r\triangle\pa_r-\frac{(2Mr_+\pa_t+a\pa_\phi)^2}
{(r-r_+)(r_+-r_-)}+\frac{(2Mr_-\pa_t+a\pa_\phi)^2}{(r-r_-)
(r_+-r_-)}+(r+2M)^2\om^2\right]R=K_lR,
\end{equation}
where $\triangle=(r-r_+)(r-r_-)$ is given previously.

\subsection{The original hidden conformal symmetry}

If we adopt the low-frequency condition
\begin{equation}\label{e:omMcon1}
 \om M\ll1,
\end{equation}
and the near-region condition
\begin{equation}\label{e:omrcon2}
 r\ll\frac{1}{\om},
\end{equation}
the first equation (\ref{e:KerrKGeq1}) reduces to
\begin{equation}\label{e:KerrKGeq11}
 \nabla^2_{S^2}S(\th,\phi)=-K_lS(\th,\phi),
\end{equation}
with $K_l=l(l+1)$. The second equation (\ref{e:KerrKGeq2}) reduces
to
\begin{equation}\label{e:KerrKGeq22}
 \left[\pa_r\triangle\pa_r-\frac{(2Mr_+\om-am)^2}
{(r-r_+)(r_+-r_-)}+\frac{(2Mr_-\om-am)^2}{(r-r_-)
(r_+-r_-)}\right]R=K_lR,
\end{equation}
whose solutions are hypergeometric functions, which are invariant
under the $SL(2,R)_L\times SL(2,R)_R$ symmetry. This indicates that
there exists some hidden conformal symmetry since there seems to be
no $AdS_3$ geometry accounting for them.

As stated in \cite{Castro:2010fd}, the ``near region" does not refer
to the near-horizon region. For small enough $\om$, the distance $r$
can be arbitrarily large. However, it should also be noticed that
the last term in Eq.\ (\ref{e:KerrKGeq2}) can not be neglected
compared with the second or the third term under the conditions
(\ref{e:omMcon1}) and (\ref{e:omrcon2}) when $|r-r_+|$ or $|r-r_-|$
is too large. Hence, the near region could not be too far from the
horizons.

\subsection{Conformal symmetry on the warped AdS$_3$ segment}

To be consistent with \cite{Castro:2010fd}, we define the
temperatures
\begin{equation}
 T_L=\frac{r_++r_-}{4\pi a}=\frac{1+\al^2}{4\pi\al},
\textrm{ }\textrm{ }\textrm{ } T_R=\frac{r_+-r_-}{4\pi
a}=\frac{1-\al^2}{4\pi\al}.
\end{equation}
Note that $2M/r_+=1+\al^2$ and $\la\ep=1-\al^2$. The temperatures
are related to the Hawking temperature through:
$T_H^{-1}=(2M^2/a)(T_L^{-1}+T_R^{-1})$.

We now consider the KG equation (\ref{e:KerrKGeq1}) and
(\ref{e:KerrKGeq2}) in the coordinate system redefined in Eqs.\
(\ref{e:Kerrred1}), (\ref{e:Kerrred2}), (\ref{e:Kerrred3}) and
(\ref{e:Kerrred4}), which lead to the near-horizon AdS$_3$ segment
(\ref{e:genKerrAdS2}) for general non-extremal Kerr. To be precise,
we here summarise the relation between the new coordinates
$(\hat{t},\hat{\phi})$ and the original ones $(t,\phi)$ defined in
these equations:
\begin{equation}
 t=2r_+\frac{T_L}{T_R}\hat{t}, \textrm{ }\textrm{ }\textrm{ }
\phi=\frac{1}{2\pi T_L}\left(\hat{\phi}
+\frac{T_L}{T_R}\hat{t}\right).
\end{equation}
\begin{eqnarray}
 \pa_t=\frac{1}{2r_+}\left(\frac{T_R}{T_L}\pa_{\hat{t}}-
\pa_{\hat{\phi}}\right), && \pa_\phi=2\pi T_L\pa_{\hat{\phi}}.
\end{eqnarray}
Instead of taking the two conditions (\ref{e:omMcon1}) and
(\ref{e:omrcon2}), we adopt the condition $r_+\om\ll1$, under which
the angular part (\ref{e:KerrKGeq1}) and the radial part
(\ref{e:KerrKGeq2}) of the KG equation then become respectively
\begin{equation}\label{e:KerrKGredeq}
 %\nabla^2_{S^2}S(\th,\phi)=
 \left[\frac{1}{\sin\th}\pa_\th\sin\th\pa_\th+\frac{4\pi^2T_L^2}
{\sin^2\th}\pa_{\hat{\phi}}^2\right]S(\th,\hat{\phi})
=-l(l+1)S(\th,\hat{\phi}),
\end{equation}
\begin{eqnarray}\label{e:KGKerrAdS}
 \frac{1}{4}\left[\frac{1}{\sinh(2\eta)}\pa_\eta\sinh
(2\eta)\pa_\eta-\frac{1}{\sinh^2\eta}\pa_{\hat{t}}^2
+\frac{1}{\cosh^2\eta}\left(\al^2\pa_{\hat{t}}+(1+\al^2)
\pa_{\hat{\phi}}\right)^2\right]R(\eta,\hat{t},\hat{\phi})
\nonumber \\
=l(l+1)R(\eta,\hat{t},\hat{\phi}),
\end{eqnarray}
where $\eta=\rho/2$, who satisfies the constraint
(\ref{e:KerrAdSrhorang}). Note that this is not exactly the KG
equation on the near-horizon geometry (\ref{e:genKerrAdS2}) or
(\ref{e:genKerrAdS3}), except in the extreme limit
$\al\rightarrow1$. This mismatch is caused by the difference in
taking near-horizon limit in the wave equation and in the Kerr
metric with the redefined coordinates $(\hat{t},\hat{\phi})$. But,
this already indicates that the near-horizon AdS segment
(\ref{e:genKerrAdS2}) is closely relevant to the wave equation
(\ref{e:KGKerrAdS}).

The equation (\ref{e:KGKerrAdS}) reminds us of the equation of a
massless scalar propagating in an $AdS_3$ space of the form:
\begin{equation}\label{e:ads3}
 ds^2=-\sinh^2\eta d\tau^2+d\eta^2+\cosh^2\eta d\si^2,
\end{equation}
in which, the Laplacian is
\begin{equation}\label{e:ads3lap}
 \nabla_{AdS_3}^2=\frac{1}{\sinh(2\eta)}\pa_\rho\sinh
(2\eta)\pa_\rho-\frac{\pa_\tau^2}{\sinh^2\eta}+\frac
{\pa_\si^2}{\cosh^2\eta}.
\end{equation}
So Eq. (\ref{e:KGKerrAdS}) can be rewritten as:
$(1/4)\nabla_{AdS_3}^2R=l(l+1)R$, with the coordinate relations:
\begin{eqnarray}
 \hat{t}=\tau+\al^2\si, &&
\hat{\phi}=(1+\al^2)\si.
\end{eqnarray}
The $AdS_3$ is a Lorentzian counterpart to $S^3$ with the isometry
group $SO(4)\simeq SU(2)\times SU(2)$. The generators for this space
are given by (similar to those given in \cite{Cvetic:1997uw})
\begin{eqnarray}\label{e:htausir}
 H_1(\tau,\si)=\frac{i}{2}e^{(\tau+\si)}[\pa_\eta-(\coth\eta
\pa_{\tau}+\tanh\eta\pa_\si)],
\nonumber \\
H_{-1}(\tau,\si)=-\frac{i}{2}e^{-(\tau+\si)}[\pa_\eta+(\coth
\eta\pa_{\tau}+\tanh\eta\pa_\si)],
\\
H_0(\tau,\si)=-\frac{i}{2}(\pa_{\tau}+\pa_\si), \nonumber
\end{eqnarray}
for the right-hand sector, and
$\bar{H}_n(\eta,\tau,\si)=H_n(\eta,\tau,-\si)$ for the left-hand
sector. The vectors $\bar{H}_n$ and $H_n$ satisfy the $SL(2,R)_L$
and $SL(2,R)_R$ algebra, respectively, with an extra
$[H_n,\bar{H}_m]=0$.

From the above equations, we know that the coordinates $(\tau,\si)$
are related to $(t,\phi)$ via
\begin{equation}
 t=2r_+\frac{T_L}{T_R}(\tau+\al^2\si),
\textrm{ }\textrm{ }\textrm{ } \phi=\frac{1}{2\pi T_R}(\tau+\si).
\end{equation}
Thus, the generators in the original coordinates $(t,\phi)$ can be
written down
\begin{eqnarray}\label{e:htphir}
 H_1(t,\phi)=ie^{2\pi T_R\phi}\left\{\pa_\rho-\frac{1}
{\sinh\rho}\left[2M\left(1+\cosh\rho\frac{T_L}{T_R}\right)
\pa_t+\frac{\cosh\rho}{2\pi T_R}\pa_\phi\right]\right\},
\nonumber \\
H_{-1}(t,\phi)=-ie^{-2\pi T_R\phi}\left\{\pa_\rho+
\frac{1}{\sinh\rho}\left[2M\left(1+\cosh\rho\frac{T_L}
{T_R}\right)\pa_t+\frac{\cosh\rho}{2\pi T_R}\pa_\phi
\right]\right\},
\\
H_0(t,\phi)=-i\left(2M\frac{T_L}{T_R}\pa_t+\frac{1}{2{\pi}T_R}
\pa_\phi\right), \nonumber
\end{eqnarray}
and
\begin{eqnarray}\label{e:htphil}
 \bar{H}_1(t,\phi)=ie^{\frac{1}{2M}t-2\pi T_L\phi}\left\{
\pa_\rho-\frac{1}{\sinh\rho}\left[2M\left(\cosh\rho
+\frac{T_L}{T_R}\right)\pa_t+\frac{1}{2\pi T_R}\pa_\phi\right]
\right\},
\nonumber \\
\bar{H}_{-1}(t,\phi)=-ie^{-\frac{1}{2M}t+2\pi T_L\phi}\left\{
\pa_\rho+\frac{1}{\sinh\rho}\left[2M\left(\cosh\rho+\frac{T_L}
{T_R}\right)\pa_t+\frac{1}{2\pi T_R} \pa_\phi\right] \right\},
\\
\bar{H}_0(t,\phi)=-2iM\pa_t. \nonumber
\end{eqnarray}
We would like to stress that the valid values of $\rho$ in the above
formulas satisfies the condition (\ref{e:KerrAdSrhorang}). If we
change back to the original radial coordinate $r$ through
$r=r_+[1+\la\ep\sinh^2(\rho/2)]$ as given in Eq.\
(\ref{e:Kerrred3}), we get the results obtained in
\cite{Castro:2010fd}. The radial wave equation can be expressed as
the $SL(2,R)$ Casimir
\begin{equation}
 H^2(t,\phi)R(\rho)=\bar{H}^2(t,\phi)R(\rho)=l(l+1)R(\rho).
\end{equation}
The $SL(2,R)_L\times SL(2,R)_R$ weights of the field $\Phi$ are
$(h_L,h_R)=(l+1,l+1)$.

Hence, we have shown that the ``hidden" conformal symmetry in the
wave equation should arise from the near-horizon $AdS_3$ segment
like (\ref{e:genKerrAdS2}) or (\ref{e:genKerrAdS3}). This $AdS$
segment can be continuously extended to the full AdS space with
$\rho\in[0,\infty)$ derived in the near-extreme case
\cite{Castro:2009jf}. If the near-horizon geometry is strictly
Rindler space, the vectors are those given in Eqs.\ (\ref{e:htphir})
and (\ref{e:htphil}) but with the replacements of
$\sinh\al\rightarrow1/\al$ and $\cosh\al\rightarrow1$, which do not
satisfy the $SL(2,R)_L\times SL(2,R)_R$ algebra.

As observed in \cite{Castro:2010fd}, the vectors are not globally
defined because they are not periodic under the periodic
identification $\phi\sim\phi+2\pi$. The $SL(2,R)_L\times SL(2,R)_R$
symmetry is spontaneously broken by the periodic $\phi$ down to the
$U(1)_L\times U(1)_R$ subgroup, which is generated by
$(\bar{H}_0,H_0)$.

The Kerr/CFT correspondence
\cite{Guica:2008mu,Matsuo:2009sj,Castro:2009jf} suggests that Kerr
black holes at extremality could be described by a two-dimensional
CFT. The central charges derived from the analysis of the asymptotic
symmetry group (ASG) are: $c_L=c_R=12J$. The entropy computed in the
dual CFT using Cardy's formula:
\begin{equation}
 S=\frac{\pi^2}{3}(c_LT_L+c_RT_R)=2\pi Mr_+,
\end{equation}
agrees with the Bekenstein-Hawking result for black hole entropy. In
\cite{Castro:2010fd}, it is also noted that the agreement is still
true for any values of $M$ and $J$. This implies that the Kerr/CFT
correspondence at extremality might be continued to the off-extreme
case, even including the Schwarzschild black holes. If so, the
$AdS_3$ segment discussed above for non-extreme Kerr black holes
might provide the geometrical explanation to this extension. But
this is hard to be verified because the AdS segment in off-extreme
case is only a portion of the AdS space in the near-extreme case.
The coordinate $\rho$ can not reach infinity and so the original ASG
approach does not apply in off-extreme case. We do not know how to
calculate the central charge to verify the entropy formular.

\subsection{Scattering}

We shall look at the scattering amplitudes on the AdS segment. It is
convenient to use the coordinate system $(\tau,\si)$, in which the
scalar can be expressed as:
$\Phi=e^{-i\omega_\tau\tau+im_\si\si}R(\eta)S(\th)$. Then the
solutions to the radial KG equation are
\begin{eqnarray}
 R(\eta)=\frac{(\coth\eta)^{i\om_\tau}}{(\cosh\eta)^{2(l+1)}}
F\left(1+l-\frac{i}{2}(\om_\tau-m_\si),1+l-
\frac{i}{2}(\om_\tau+m_\si);1-i\om_\tau; \tanh^2\eta\right).
\end{eqnarray}

The hypergeometric function $F(\al,\be,\ga,z=\tanh^2\eta)$ can be
transformed into the form $F(\al,\be,\ga,1-z=\cosh^{-2}\eta)$. In
order to match the boundary conditions of the near and far regions,
we need to know the solution at large $\eta$. We can expand around
the pole $1-z\sim0$ to leading order:
\begin{equation}
 R(\eta)\sim A(\cosh\eta)^{2l}+B(\cosh\eta)^{-2(l+1)}.
\end{equation}
where
\begin{equation}
 A=\frac{\Ga(1-i\om_\tau)\Ga(1+2l)}{\Ga(1+l-\frac{i}{2}(\om_\tau
-m_\si))\Ga(1+l-\frac{i}{2}(\om_\tau+m_\si))},
\end{equation}
\begin{equation}
 B=\frac{\Ga(1-i\om_\tau)\Ga(-1-2l)}{\Ga(-l-\frac{i}{2}(\om_\tau
+m_\si))\Ga(-l-\frac{i}{2}(\om_\tau-m_\si))}.
\end{equation}

Since $\cosh^{-2}\eta$ drops to zero rapidly with $\eta$ increasing,
this approximate solution can be trusted as the black holes are not
far away from extremality. However, for the general off-extreme
case, the maximum value of $\eta$ may not be large enough. So we
should consider higher order terms in the expanded solution, as
expected. This will be left for further study. In the original paper
\cite{Castro:2010fd}, the leading term is adopted as the solution
for general black holes because they think that $r$ can be
arbitrarily large by choosing small enough $\om$. But, as we have
argued above, $r$ can not be too far away from the horizon to
guarantee that we can pass from Eq.\ (\ref{e:KerrKGeq2}) to the
approximate equation (\ref{e:KerrKGeq22}).

In the near-extreme case, the absorption cross section is then
proportional to
\begin{equation}
 P_{\textrm{abs}}\sim|A|^{-2}\sim\sinh\om_\tau\left|\Ga
\left(h_L-\frac{i}{2}(\om_\tau-m_\si)\right)
\right|^2\left|\Ga\left(h_R-\frac{i}{2}(\om_\tau+m_\si)
\right)\right|^2.
\end{equation}
Identifying $\om_L=\pi T_L(\om_\tau-m_\si)$ and $\om_R=\pi
T_R(\om_\tau+m_\si)$, we can get the standard cross-section for 2D
CFT given in \cite{Castro:2010fd}.

%%%%%%%%%%%%%%%%%%%%%%%%%%%%%%%%%%%%%%%%%%%%%%%%%%%%%%%%%%%%%%%%%%%%%%%%%%%%
\section{Conclusions}
\label{sec:conclusion}
%%%%%%%%%%%%%%%%%%%%%%%%%%%%%%%%%%%%%%%%%%%%%%%%%%%%%%%%%%%%%%%%%%%%%%%%%%%%

It is shown that hidden conformal symmetries near horizons are
derived because the near-horizon geometries of non-extreme black
holes are essentially taken as some AdS segments instead of simply
Rindler space, though the former contain the Rindler wedge as well
in the neighborhood of the horizon. These AdS segments can be viewed
as the remnants of the AdS spaces in near-extreme case. That is, the
AdS spaces in near-extreme case do not disappear immediately and a
portion of them (the part close to the horizon) remain in the case
away from extremality. Since these AdS segments do not contain the
boundary at spatial infinity, we do not know how to calculate the
central charge using the ASG method. So the suggested entropy
formula can not be easily verified. The absorption possibility is
also not able to be computed using the ordinary method in this
situation. This may demand higher order corrections to be taken into
account, which is the target of future work.

\section*{Acknowledgements\markboth{Acknowledgements}{Acknowledgements}}

\noindent We would like to thank Chiang-Mei Chen and Zhiguang Xiao
for useful discussions.

\appendix

%\section*{Appendices}
\section*{Appendix}

\setcounter{equation}{0}

\section{AdS segment in 5D rotating black holes}

The near-horizon AdS segments can also be obtained in higher
dimensional black holes. Here we derive the form for 5D rotating
black holes. The hidden conformal symmetry in this kind of black
holes has been discussed in
\cite{Krishnan:2010pv,Chen:2011kt,Cvetic:2011hp}. The metric is
\begin{eqnarray}\label{e:5DKerr}
 ds^2=-\frac{\triangle}{r^2\chi^2}(dt-a\sin^2\th d\phi-b\cos^2\th
d\psi)^2+\frac{r^2\chi^2}{\triangle}dr^2+\chi^2d\th^2 \nonumber \\
+\frac{\sin^2\th} {\chi^2}[adt-(r^2+a^2)d\phi]^2+\frac{\cos^2\th}
{\chi^2}[bdt-(r^2+b^2)d\psi]^2 \\
+\frac{1}{r^2\chi^2}[abdt-b(r^2+a^2)\sin^2\th
d\phi-a(r^2+b^2)\cos^2\th d\psi]^2, \nonumber
\end{eqnarray}
where
\begin{equation}
 \triangle=(r^2+a^2)(r^2+b^2)-2Mr^2, \textrm{ }\textrm{ }\textrm{ }
 \chi^2=r^2+a^2\cos^2\th+b^2\sin^2\th.
\end{equation}
The radii of the inner and outer horizons are
\begin{equation}
 r_\pm^2=M-\frac{1}{2}(a^2+b^2)\pm\{[M-\frac{1}{2}(a^2+b^2)]^2
-a^2b^2\}^{\frac{1}{2}}.
\end{equation}
The extreme limit is at $r_+=r_-=\sqrt{ab}$, with $M=(a+b)^2/2$.

It is easy to derive a similar form of the near-horizon geometry to
the one (\ref{e:genKerrAdS1}) or (\ref{e:genKerrAdS2}) for 4D Kerr.
In what follows, we only present the form corresponding to Eq.\
(\ref{e:genKerrAdS3}).

We adopt the following notations and coordinate redefinitions:
\begin{equation}\label{e:}
 r^2=r_+^2(1+\la U), \textrm{ }\textrm{ }\textrm{ }
t=\frac{r_+\bar{t}}{\la},
\end{equation}
\begin{equation}\label{e:}
 \phi=\frac{\al\bar{\phi}}{1+\al^2}+\frac{2\al\bar{t}}{A},
\textrm{ }\textrm{ }\textrm{ }
\psi=\frac{\be\bar{\psi}}{1+\be^2}+\frac{2\be\bar{t}}{B},
\end{equation}
\begin{equation}\label{e:}
 A=1+\al^2\be^2+2\al^2, \textrm{ }\textrm{ }\textrm{ }
B=1+\al^2\be^2+2\be^2,
\end{equation}
\begin{equation}\label{e:}
 \al=\frac{a}{r_+},  \textrm{ }\textrm{ }\textrm{ }
\be=\frac{b}{r_+}, \textrm{ }\textrm{ }\textrm{ }
\la\ep=\frac{r_+^2-r_-^2}{r_+^2}=1-\al^2\be^2.
\end{equation}
The last relation comes from $r_+^2r_-^2=a^2b^2$. In this coordinate
system, the Hawking temperature and Bekenstein-Hawking entropy are
expressed respectively:
\begin{equation}\label{e:}
 T_H=\frac{r_+(1-\al^2\be^2)}{2\pi(1+\al^2)(1+\be^2)},
\textrm{ }\textrm{ }\textrm{ }
S_{BH}=\frac{\pi^2r_+}{2}(1+\al^2)(1+\be^2).
\end{equation}

Taking the near-horizon limit $\la\ll1$ and $\la U\ll1$, the metric
(\ref{e:5DKerr}) in the new coordinates is
\begin{eqnarray}
 ds^2=r_+^2\Ga\left\{-\frac{\La^2U(U+\ep)}{16\Ga^2}d\bar{t}^2+
\frac{1}{4U(U+\ep)}dU^2+d\th^2+\frac{\al^2\sin^2\th}{A^2\Ga^2}\left[(2U+\ep)d\bar{t}+
Ad\bar{\phi}\right]^2 \right.
\nonumber \\
\left. +\frac{\be^2\cos^2\th}{B^2\Ga^2}\left[(2U+\ep)d\bar{t}+
Bd\bar{\psi}\right]^2+\al^2\be^2\left[\frac{\La(2U+\ep)}{4\Ga(1+\al^2\be^2)}d\bar{t}
+\sin^2\th d\bar{\phi} +\cos^2\th d\bar{\psi}\right]^2\right\},
\end{eqnarray}
where
\begin{equation}\label{e:}
 \Ga(\th)=1+\al^2\cos^2\th+\be^2\sin^2\th,
\end{equation}
\begin{equation}\label{e:}
 \La(\th)=\frac{4(1+\al^2\be^2)[1+\al^2\be^2+2(\al^2\cos^2\th
+\be^2\sin^2\th)]}{(1+\al^2\be^2+2\al^2)(1+\al^2\be^2+2\be^2)}.
\end{equation}
The condition $\la\gg1$ requires that the parameter $\ep$ satisfies
$\ep\gg1-\al^2\be^2$. In the extremal limit $\al^2\be^2=1$, we have
$\La=6\Ga/(1+\al^2)(1+\be^2)$. For identical spin $a=b$, we have
$\La=\Ga$.

Let us further redefine
\begin{equation}\label{e:}
 U=\ep\sinh^2\rho,  \textrm{ }\textrm{ }\textrm{ }
\bar{t}=\frac{\hat{t}}{\ep},
\end{equation}
The metric becomes finally
\begin{eqnarray}
 ds^2=r_+^2\Ga\left\{-\frac{\La^2}{64\Ga^2}\sinh^2(2\rho)
d\hat{t}^2+d\rho^2+d\th^2+\frac{\al^2\sin^2\th}{A^2\Ga^2}
\left[\cosh(2\rho)d\hat{t}+Ad\bar{\phi}\right]^2 \right.
\nonumber \\
+\frac{\be^2\cos^2\th}{B^2\Ga^2}\left[\cosh(2\rho)d\hat{t}
+Bd\bar{\psi}\right]^2+\al^2\be^2\left[\frac{\La\cosh(2\rho)}
{4\Ga(1+\al^2\be^2)}d\hat{t}\right.
\\
\left.\left.+\sin^2\th d\bar{\phi}+\cos^2\th d\bar{\psi}
\right]^2\right\}, \nonumber
\end{eqnarray}
where
\begin{equation}
 \rho\sim\sinh^{-1}\left(\frac{1}{\sqrt{\ep}}\right)
\ll\sinh^{-1}\left(\frac{1}{\sqrt{1-\al^2\be^2}}\right).
\end{equation}
The coordinate $\rho$ is constrained because we have set
$U\sim\mathcal{O}(1)$, as done in the text. Towards the extreme
limit, we have the AdS space with $\rho\in[0,\infty)$.

%\newpage
\bibliographystyle{JHEP}
\bibliography{b}

\providecommand{\href}[2]{#2}\begingroup\raggedright\begin{thebibliography}{10}

\bibitem{Bardeen:1999px}
J.~M. Bardeen and G.~T. Horowitz, {\it {The extreme Kerr throat geometry: A
  vacuum analog of AdS(2) x S(2)}},  {\em Phys. Rev.} {\bf D60} (1999) 104030,
  [\href{http://xxx.lanl.gov/abs/hep-th/9905099}{{\tt hep-th/9905099}}].

\bibitem{Guica:2008mu}
M.~Guica, T.~Hartman, W.~Song, and A.~Strominger, {\it {The Kerr/CFT
  Correspondence}},  {\em Phys. Rev.} {\bf D80} (2009) 124008,
  [\href{http://xxx.lanl.gov/abs/0809.4266}{{\tt arXiv:0809.4266}}].

\bibitem{Bredberg:2011hp}
I.~Bredberg, C.~Keeler, V.~Lysov, and A.~Strominger, {\it {Cargese Lectures on
  the Kerr/CFT Correspondence}},  \href{http://xxx.lanl.gov/abs/1103.2355}{{\tt
  arXiv:1103.2355}}.

\bibitem{Matsuo:2009sj}
Y.~Matsuo, T.~Tsukioka, and C.-M. Yoo, {\it {Another Realization of Kerr/CFT
  Correspondence}},  {\em Nucl. Phys.} {\bf B825} (2010) 231--241,
  [\href{http://xxx.lanl.gov/abs/0907.0303}{{\tt arXiv:0907.0303}}].

\bibitem{Rasmussen:2009ix}
J.~Rasmussen, {\it {Isometry-preserving boundary conditions in the Kerr/CFT
  correspondence}},  {\em Int. J. Mod. Phys.} {\bf A25} (2010) 1597--1613,
  [\href{http://xxx.lanl.gov/abs/0908.0184}{{\tt arXiv:0908.0184}}].

\bibitem{Castro:2009jf}
A.~Castro and F.~Larsen, {\it {Near Extremal Kerr Entropy from $AdS_2$ Quantum
  Gravity}},  {\em JHEP} {\bf 12} (2009) 037,
  [\href{http://xxx.lanl.gov/abs/0908.1121}{{\tt arXiv:0908.1121}}].

\bibitem{Maldacena:1998uz}
J.~M. Maldacena, J.~Michelson, and A.~Strominger, {\it {Anti-de Sitter
  fragmentation}},  {\em JHEP} {\bf 02} (1999) 011,
  [\href{http://xxx.lanl.gov/abs/hep-th/9812073}{{\tt hep-th/9812073}}].

\bibitem{Castro:2010fd}
A.~Castro, A.~Maloney, and A.~Strominger, {\it {Hidden Conformal Symmetry of
  the Kerr Black Hole}},  {\em Phys. Rev.} {\bf D82} (2010) 024008,
  [\href{http://xxx.lanl.gov/abs/1004.0996}{{\tt arXiv:1004.0996}}].

\bibitem{Chen:2010as}
C.-M. Chen and J.-R. Sun, {\it {Hidden Conformal Symmetry of the
  Reissner-Nordstr{\o}m Black Holes}},  {\em JHEP} {\bf 08} (2010) 034,
  [\href{http://xxx.lanl.gov/abs/1004.3963}{{\tt arXiv:1004.3963}}].

\bibitem{Chen:2010yu}
C.-M. Chen, Y.-M. Huang, J.-R. Sun, M.-F. Wu, and S.-J. Zou, {\it {On
  Holographic Dual of the Dyonic Reissner-Nordstr\'om Black Hole}},  {\em Phys.
  Rev.} {\bf D82} (2010) 066003, [\href{http://xxx.lanl.gov/abs/1006.4092}{{\tt
  arXiv:1006.4092}}].

\bibitem{Wang:2010qv}
Y.-Q. Wang and Y.-X. Liu, {\it {Hidden Conformal Symmetry of the Kerr-Newman
  Black Hole}},  {\em JHEP} {\bf 08} (2010) 087,
  [\href{http://xxx.lanl.gov/abs/1004.4661}{{\tt arXiv:1004.4661}}].

\bibitem{Chen:2010xu}
B.~Chen and J.~Long, {\it {Real-time Correlators and Hidden Conformal Symmetry
  in Kerr/CFT Correspondence}},  {\em JHEP} {\bf 06} (2010) 018,
  [\href{http://xxx.lanl.gov/abs/1004.5039}{{\tt arXiv:1004.5039}}].

\bibitem{Chen:2010zwa}
D.~Chen, P.~Wang, and H.~Wu, {\it {Hidden conformal symmetry of rotating
  charged black holes}},  {\em Gen. Rel. Grav.} {\bf 43} (2011) 181--190,
  [\href{http://xxx.lanl.gov/abs/1005.1404}{{\tt arXiv:1005.1404}}].

\bibitem{Wang:2010ic}
H.~Wang, D.~Chen, B.~Mu, and H.~Wu, {\it {Hidden conformal symmetry of extreme
  and non-extreme Einstein-Maxwell-Dilaton-Axion black holes}},  {\em JHEP}
  {\bf 11} (2010) 002, [\href{http://xxx.lanl.gov/abs/1006.0439}{{\tt
  arXiv:1006.0439}}].

\bibitem{Shao:2010cf}
K.-N. Shao and Z.~Zhang, {\it {Hidden Conformal Symmetry of Rotating Black Hole
  with four Charges}},  {\em Phys. Rev.} {\bf D83} (2011) 106008,
  [\href{http://xxx.lanl.gov/abs/1008.0585}{{\tt arXiv:1008.0585}}].

\bibitem{Setare:2010sy}
M.~R. Setare and V.~Kamali, {\it {Hidden Conformal Symmetry of Rotating Black
  Holes in Minimal Five-Dimensional Gauged Supergravity}},  {\em Phys. Rev.}
  {\bf D82} (2010) 086005, [\href{http://xxx.lanl.gov/abs/1008.1123}{{\tt
  arXiv:1008.1123}}].

\bibitem{Ghezelbash:2010vt}
A.~M. Ghezelbash, V.~Kamali, and M.~R. Setare, {\it {Hidden Conformal Symmetry
  of Kerr-Bolt Spacetimes}},  {\em Phys. Rev.} {\bf D82} (2010) 124051,
  [\href{http://xxx.lanl.gov/abs/1008.2189}{{\tt arXiv:1008.2189}}].

\bibitem{Huang:2010yg}
Y.-C. Huang and F.-F. Yuan, {\it {Hidden conformal symmetry of extremal
  Kaluza-Klein black hole in four dimensions}},  {\em JHEP} {\bf 03} (2011)
  029, [\href{http://xxx.lanl.gov/abs/1012.5453}{{\tt arXiv:1012.5453}}].

\bibitem{Peng:2011zza}
J.-J. Peng and S.-Q. Wu, {\it {Hidden conformal symmetry of rotating black
  holes in the five-dimensional Goedel universe}},  {\em Gen.Rel.Grav.} {\bf
  43} (2011) 2743--2756.

\bibitem{Chen:2011kt}
B.~Chen and J.-j. Zhang, {\it {General Hidden Conformal Symmetry of 4D
  Kerr-Newman and 5D Kerr Black Holes}},
  \href{http://xxx.lanl.gov/abs/1107.0543}{{\tt arXiv:1107.0543}}.

\bibitem{Bertini:2011ga}
S.~Bertini, S.~L. Cacciatori, and D.~Klemm, {\it {Conformal structure of the
  Schwarzschild black hole}},  \href{http://xxx.lanl.gov/abs/1106.0999}{{\tt
  arXiv:1106.0999}}.

\bibitem{Carlip:1998wz}
S.~Carlip, {\it {Black hole entropy from conformal field theory in any
  dimension}},  {\em Phys. Rev. Lett.} {\bf 82} (1999) 2828--2831,
  [\href{http://xxx.lanl.gov/abs/hep-th/9812013}{{\tt hep-th/9812013}}].

\bibitem{Solodukhin:1998tc}
S.~N. Solodukhin, {\it {Conformal description of horizon's states}},  {\em
  Phys. Lett.} {\bf B454} (1999) 213--222,
  [\href{http://xxx.lanl.gov/abs/hep-th/9812056}{{\tt hep-th/9812056}}].

\bibitem{Sachs:2001qb}
I.~Sachs and S.~N. Solodukhin, {\it {Horizon holography}},  {\em Phys. Rev.}
  {\bf D64} (2001) 124023, [\href{http://xxx.lanl.gov/abs/hep-th/0107173}{{\tt
  hep-th/0107173}}].

\bibitem{Govindarajan:2000ag}
T.~R. Govindarajan, V.~Suneeta, and S.~Vaidya, {\it {Horizon states for AdS
  black holes}},  {\em Nucl. Phys.} {\bf B583} (2000) 291--303,
  [\href{http://xxx.lanl.gov/abs/hep-th/0002036}{{\tt hep-th/0002036}}].

\bibitem{Birmingham:2001qa}
D.~Birmingham, K.~S. Gupta, and S.~Sen, {\it {Near-horizon conformal structure
  of black holes}},  {\em Phys. Lett.} {\bf B505} (2001) 191--196,
  [\href{http://xxx.lanl.gov/abs/hep-th/0102051}{{\tt hep-th/0102051}}].

\bibitem{Gupta:2001bg}
K.~S. Gupta and S.~Sen, {\it {Further evidence for the conformal structure of a
  Schwarzschild black hole in an algebraic approach}},  {\em Phys. Lett.} {\bf
  B526} (2002) 121--126, [\href{http://xxx.lanl.gov/abs/hep-th/0112041}{{\tt
  hep-th/0112041}}].

\bibitem{deAlfaro:1976je}
V.~de~Alfaro, S.~Fubini, and G.~Furlan, {\it {Conformal Invariance in Quantum
  Mechanics}},  {\em Nuovo Cim.} {\bf A34} (1976) 569.

\bibitem{Matsuo:2010in}
Y.~Matsuo, T.~Tsukioka, and C.-M. Yoo, {\it {Notes on the Hidden Conformal
  Symmetry in the Near Horizon Geometry of the Kerr Black Hole}},  {\em Nucl.
  Phys.} {\bf B844} (2011) 146--163,
  [\href{http://xxx.lanl.gov/abs/1007.3634}{{\tt arXiv:1007.3634}}].

\bibitem{Cvetic:2011hp}
M.~Cvetic and F.~Larsen, {\it {Conformal Symmetry for General Black Holes}},
  \href{http://xxx.lanl.gov/abs/1106.3341}{{\tt arXiv:1106.3341}}.

\bibitem{Franzin:2011wi}
E.~Franzin and I.~Smolic, {\it {A new look at hidden conformal symmetries of
  black holes}},  {\em JHEP} {\bf 09} (2011) 081,
  [\href{http://xxx.lanl.gov/abs/1107.2756}{{\tt arXiv:1107.2756}}].

\bibitem{Carroll:2009maa}
S.~M. Carroll, M.~C. Johnson, and L.~Randall, {\it {Extremal limits and black
  hole entropy}},  {\em JHEP} {\bf 11} (2009) 109,
  [\href{http://xxx.lanl.gov/abs/0901.0931}{{\tt arXiv:0901.0931}}].

\bibitem{Garousi:2009zx}
M.~R. Garousi and A.~Ghodsi, {\it {The RN/CFT Correspondence}},  {\em Phys.
  Lett.} {\bf B687} (2010) 79--83,
  [\href{http://xxx.lanl.gov/abs/0902.4387}{{\tt arXiv:0902.4387}}].

\bibitem{Lowe:2011wu}
D.~A. Lowe, I.~Messamah, and A.~Skanata, {\it {Scaling dimensions in hidden
  Kerr/CFT}},  \href{http://xxx.lanl.gov/abs/1105.2035}{{\tt arXiv:1105.2035}}.

\bibitem{Cvetic:1997uw}
M.~Cvetic and F.~Larsen, {\it {General rotating black holes in string theory:
  Greybody factors and event horizons}},  {\em Phys. Rev.} {\bf D56} (1997)
  4994--5007, [\href{http://xxx.lanl.gov/abs/hep-th/9705192}{{\tt
  hep-th/9705192}}].

\bibitem{Krishnan:2010pv}
C.~Krishnan, {\it {Hidden Conformal Symmetries of Five-Dimensional Black
  Holes}},  {\em JHEP} {\bf 07} (2010) 039,
  [\href{http://xxx.lanl.gov/abs/1004.3537}{{\tt arXiv:1004.3537}}].

\end{thebibliography}\endgroup

\end{document}